\documentclass[a4paper,12pt]{article}
\usepackage[a4paper,text={16.8cm,22.4cm}]{geometry}
\usepackage{amsmath,amssymb,bm,epsfig,psfrag}
\usepackage{cite}

\allowdisplaybreaks 
\newcommand{\ff}{f\hspace{-0.4em}f}

\begin{document}

\begin{titlepage}

\begin{flushright}
  MZ-TH/09-48\\[2mm]
  April 12, 2009
\end{flushright}

\vspace{5ex}

\begin{center}
  \textbf{\Large Threshold expansion at order $\bm{\alpha_s^4}$ for the $\bm{t\bar{t}}$
    invariant mass distribution at hadron colliders}

  \vspace{7ex}

  \textsc{ Valentin Ahrens, Andrea Ferroglia, Matthias Neubert,
    \\
    Ben D.~Pecjak, and Li Lin Yang}

  \vspace{2ex}

  \textsl{Institut f\"ur Physik (THEP), Johannes Gutenberg-Universit\"at
    \\
    D-55099 Mainz, Germany}
\end{center}

\vspace{4ex}

\begin{abstract}\noindent
  We calculate the leading $\mathcal{O}(\alpha_s^4)$ contributions to the invariant mass
  distribution of top-quark pairs produced at the Tevatron and LHC, in the limit where the
  invariant mass of the $t\bar{t}$ pair approaches the partonic center-of-mass energy. Our
  results determine at NNLO in $\alpha_s$ the coefficients of all singular plus
  distributions and scale-dependent logarithms in the differential partonic cross sections
  for $q\bar{q},gg \to t\bar{t}+X$. A numerical analysis showing the effects of the NNLO
  corrections on the central values and scale dependence of the invariant mass
  distribution is performed. The NNLO corrections are found to significantly enhance the
  cross section and reduce the perturbative uncertainties compared to the NLO calculation.
\end{abstract}

\end{titlepage}

\section{Introduction}

The top quark is the heaviest known particle in the Standard Model (SM) of fundamental
interactions. Because of its large mass, it is expected to couple strongly with the fields
responsible for electroweak symmetry breaking, and the detailed study of top-quark
properties is likely to play a key role in elucidating the origin of particle masses. In
addition to observables such as the top-quark mass and total inclusive top-quark pair
production cross section, differential cross sections are also of interest. For instance,
the $t\bar{t}$ invariant mass distribution can be used to measure $m_t$
\cite{Frederix:2007gi}, and in searches for physics beyond the SM. The presence of bumps
in the smoothly decreasing $t\bar{t}$ invariant mass distribution would be a clear signal
of an $s$-channel heavy resonance \cite{Barger:2006hm}, which is predicted in many
beyond-the-SM scenarios \cite{Baur:2008uv}.

To date, thousands of top-quark events have been observed by two different experiments at
the Fermilab Tevatron, and the top-quark mass has been extracted at the percent level.
Searches for narrow width particles decaying into top-quark pairs have been pursued
\cite{:2007dz}, and results for the top-pair invariant mass distribution were recently
obtained from data collected by the CDF collaboration \cite{Aaltonen:2009iz}. The
experiments at the LHC are expected to observe millions of top-quark events per year
already in the initial low-luminosity phase, bringing the study of top-quark properties
into the realm of precision physics. In particular, the total inclusive top-quark pair
production cross section is expected to be measured with a relative error of $5\%$ to
$10\%$ at the LHC \cite{Bernreuther:2008ju}. To make optimal use of the data requires
equally precise theoretical predictions for the measured observables.

Theoretical calculations in QCD rely on the factorization formula for the differential
cross section, which is of the form
\begin{equation}
  \label{eq:vaguefact}
  d\sigma = \sum C_{ij}\otimes f_i\otimes f_j \, ,
\end{equation}
where the symbol $\otimes$ stands for a convolution. The hard-scattering kernels $C_{ij}$
are related to the partonic cross section and can be calculated as a series in $\alpha_s$,
whereas the $f_i$ are parton distribution functions (PDFs) for the partons $i=q,\bar{q},g$
in the incoming hadrons and must be extracted from data. Current theoretical predictions
are based on next-to-leading order (NLO) calculations of the total cross section
\cite{Nason:1987xz, Beenakker:1988bq, Beenakker:1990maa, Czakon:2008ii} and differential
distributions \cite{Nason:1989zy, Mangano:1991jk} in fixed-order perturbation theory,
supplemented with soft gluon resummation to next-to-leading-logarithmic (NLL) order
\cite{Kidonakis:1997gm, Bonciani:1998vc, Kidonakis:2001nj, Kidonakis:2003qe,
  Czakon:2008cx}. The NLO computations suffer from theory uncertainties larger than
$10\%$, both for Tevatron and LHC center-of-mass energies. These uncertainties are due to
our imperfect knowledge of the parton distribution functions, and also to the truncation
of the perturbative series in the strong coupling constant, which introduces a dependence
on the unphysical renormalization and factorization scales into physical predictions. This
theoretical uncertainty is typically reduced by including more terms in the perturbative
series, and for this reason the calculation of the differential partonic cross section to
next-to-next-to-leading order (NNLO) has been an area of active research
\cite{Czakon:2007ej, Czakon:2007wk, Czakon:2008zk, Bonciani:2008az, Bonciani:2009nb}.
However, due to the complexity of the diagrammatic calculations, complete NNLO results are
not yet available. In the absence of full results, there has been much recent activity in
NNLO calculations and next-to-next-to-leading-logarithmic (NNLL) resummation for the total
cross section near production threshold \cite{Moch:2008qy, Langenfeld:2009wd,
  Czakon:2009zw, Beneke:2009rj, Beneke:2009ye}, but the analogous results for differential
cross sections remain unknown.

The goal of this Letter is to present a subset of the NNLO corrections to the
hard-scattering kernels for the $t\bar{t}$ invariant mass distribution at hadron
colliders. In particular, we focus on the region where the invariant mass $M$ of the
top-quark pair approaches the partonic center-of-mass energy $\sqrt{\hat{s}}$, and compute
NNLO corrections to the hard-scattering kernels for the $q\bar{q}$ and $gg$ channels at
leading order in $1-z$, where $z=M^2/\hat{s} \to 1$ in the threshold region. The
leading-order term in this threshold expansion for the differential cross section is
equivalent to the virtual-soft approximation, and is written in terms of singular plus
distributions and delta functions in the variable $1-z$. Our results determine the
coefficients of all plus distributions of the form $[\ln^n(1-z)/(1-z)]_+$, as well as all
$\mu$-dependent pieces multiplying the $\delta(1-z)$ term, whereas a remaining piece of
the delta-function coefficient is left undetermined. The basis for our calculations is a
factorization formula for the hard-scattering kernels in the partonic threshold region,
and is described in Section \ref{sec:NNLO}. In this region, the hard-scattering kernels
factorize into products of hard functions, related to virtual corrections, and soft
functions, related to real emission in the soft limit \cite{Kidonakis:1997gm}. These
functions satisfy certain renormalization-group equations, which determine their
dependence on the scale $\mu$. By calculating the hard and soft functions at one-loop
order, and using results for the two-loop anomalous dimensions recently derived in
\cite{Mitov:2009sv, Becher:2009kw, Ferroglia:2009ep, Ferroglia:2009ii} (see also
\cite{Aybat:2006wq, Aybat:2006mz, Becher:2009cu, Becher:2009qa, Gardi:2009qi}), the
$\mu$-dependent logarithms in the hard and soft functions can be determined exactly to
NNLO using the renormalization group. The logarithms in the soft function are of the form
$\ln(\hat{s}(1-z)^2/\mu^2)$, and uniquely determine the coefficients of the
$[\ln^n(1-z)/(1-z)]_+$ distributions. This is similar in spirit to the calculations of
\cite{Kidonakis:2003qe} for the soft corrections to the NNLO differential cross section
using threshold resummation techniques in Mellin space, but goes beyond those results by
completely determining the coefficient of the $[1/(1-z)]_+$ distribution, which is
sensitive to process-dependent two-loop anomalous dimensions. In Section \ref{sec:pheno}
we perform a short numerical analysis illustrating the impact of the NNLO corrections on
the central values and scale dependence of the $t\bar{t}$ invariant mass distribution.

\section{NNLO corrections  at threshold}
\label{sec:NNLO}

In this section we discuss our method for determining NNLO corrections to the differential
$pp\,(p\bar{p}) \to t\bar{t}+X$ cross section in the threshold region. To describe the
threshold region we introduce the variables
\begin{align}
  z = \frac{M^2}{\hat s} \, , \qquad \tau = \frac{M^2}{s} \, , \qquad \rho =
  \frac{4m_t^2}{M^2} \, ,
\end{align}
where $M$ is the invariant mass of the $t\bar{t}$ pair, $\sqrt{s}$ is the hadronic
center-of-mass energy, and $\sqrt{\hat{s}}$ is the partonic center-of-mass energy. We
define the threshold region as the limit $z \to 1$, with $\rho=4m_t^2/M^2$ a generic
$\mathcal{O}(1)$ variable, which is however not too close to unity. Note that this is
different from the threshold limit $\rho \to 1$ often used for the total cross section.

The calculation of the differential cross section in the threshold region is greatly
simplified, since there is little phase space available for real gluon emission. Hard
emissions are suppressed by powers of $1-z$, and the partonic scattering process is
dominated by hard virtual corrections and the real emission of soft partons, which can be
calculated in the eikonal approximation and exponentiate into Wilson lines. The phase
space for these processes is effectively two-body, and the fully differential cross
section can be written in terms of the kinematic variables entering the Born-level result.

At the Born level the differential cross section receives contributions from the
quark-antiquark annihilation and gluon fusion channels
\begin{equation}
  \begin{aligned}
    q(p_1) + \bar{q}(p_2) &\to t(p_3) + \bar{t}(p_4) \, ,
    \\
    g(p_1) + g(p_2) &\to t(p_3) + \bar{t}(p_4) \, .
  \end{aligned}
\end{equation}
The partonic cross section is a function of the kinematic invariants
\begin{align}
  \label{eq:Mandelstam}
  \hat s=(p_1+p_2)^2 \, , \qquad t_1 = (p_1-p_3)^2-m_t^2 \, , \qquad u_1 = (p_1-p_4)^2
  -m_t^2 \, ,
\end{align}
and momentum conservation implies $\hat{s}+t_1+u_1=0$. The hadronic cross section is
obtained from the partonic one by convoluting with PDFs. The fully differential cross
section involves three variables, which can be chosen, for instance, as the invariant mass
and rapidity of the $t\bar{t}$ pair, and the scattering angle $\theta$ between $\vec{p}_1$
and $\vec{p}_3$ in the partonic center-of-mass frame. The kinematic invariants in
(\ref{eq:Mandelstam}) can be written in terms of these variables according to
\begin{align}
  \hat{s} = \frac{M^2}{z} \, , \qquad t_1 = -\frac{M^2}{2} \, (1-\beta\cos\theta) \, ,
  \qquad \beta = \sqrt{1-\rho} \, .
\end{align}
Less differential results are derived by integrating the triply differential rate over the
appropriate phase space.

In this Letter we will be interested in the invariant mass distribution of the $t\bar{t}$
pair. We shall write this distribution at threshold as
\begin{align}\label{eq:sigM}
  \frac{d\sigma^{\text{thres}}}{dM} = \frac{8\pi\beta}{3sM} \int_\tau^1 \! \frac{dz}{z}
  \left[ \ff_{gg} \left( \frac{\tau}{z} , \mu \right) C_{gg}(z,M,m_t,\mu) + \ff_{q\bar{q}}
    \left( \frac{\tau}{z} , \mu \right) C_{q\bar{q}}(z,M,m_t,\mu) \right] +
  \mathcal{O}(1-z) \, ,
\end{align}
where we have introduced the parton luminosity functions
\begin{equation}
  \ff_{ij}(y,\mu) = \int_y^1 \! \frac{dx}{x} \, f_i(x,\mu) \, f_j(y/x,\mu) \, .
\end{equation}
For simplicity, we set the factorization and renormalization scales equal and refer to
them as $\mu$. The luminosities for $q\bar{q}$ are understood to be summed over all
species of light quarks, and we have made explicit that mixed channels such as $q g$ are
power suppressed by $1-z$ in the threshold region \cite{Kidonakis:1997gm}. The
hard-scattering kernels $C_{ij}$ factorize into hard functions, related to virtual
corrections, and soft functions, related to soft real emissions \cite{Kidonakis:1997gm}.
The result is
\begin{align}\label{eq:FF}
  C_i(z,M,m_t,\mu) = \int_{-1}^1 \! d\cos\theta \, {\rm Tr} \Big[
  \bm{H}_i(M,m_t,\cos\theta,\mu) \, \bm{S}_i(\sqrt{\hat{s}}(1-z),M,m_t,\cos\theta,\mu)
  \Big] \, .
\end{align}
We have used a label $i=q\bar{q},gg$ to distinguish the different partonic channels, but
will suppress it in the formulas below, referring instead to a single coefficient
function. Note that the hard and soft functions are matrices in a color decomposition of
the QCD amplitudes. We have used $\hat{s}=M^2/z \approx M^2$ everywhere except in the
first argument of the soft function, since we expect that the exact results contain
logarithms of the form $\ln(\hat{s}(1-z)^2/\mu^2)$, as in the case of Drell-Yan production
\cite{Becher:2007ty}. In \cite{Ahrens:2010zv}, we will give details on the derivation of
the factorization formula (\ref{eq:FF}) in soft-collinear effective theory.

The perturbative expansion of the hard-scattering kernel can be written as
\begin{align}\label{eq:Cexp}
  C = \alpha_s^2 \left[ C^{(0)} + \frac{\alpha_s}{4\pi}\,C^{(1)} + \left(
      \frac{\alpha_s}{4\pi} \right)^2 C^{(2)} + \cdots \right] .
\end{align}
The purpose of this Letter is to present new results for the NNLO coefficient $C^{(2)}$. As
mentioned in the introduction, our derivation relies on the renormalization-group
equations for the hard and soft functions. The soft function contains singular plus
distributions, and its renormalization-group equation is non-local. It is convenient to
avoid this complication by working instead with the Laplace transform of the soft function
\cite{Becher:2006nr, Becher:2006mr}, defined as
\begin{align}\label{eq:stilde}
  \bm{\widetilde{s}}(L,M,m_t,\cos\theta,\mu) = \int_0^\infty \! d\omega \,
  e^{-\sigma\omega} \bm{S}(\omega,M,m_t,\cos\theta,\mu) \, , \qquad \sigma =
  \frac{1}{e^{\gamma_E}\mu e^{L/2}} \, .
\end{align}
Whereas the soft function contains distributions, the Laplace-transformed function is a
polynomial in its first argument and satisfies a local evolution equation. To rewrite the
hard-scattering kernel in terms of this function, we introduce the notation
\begin{align}\label{eq:CalC}
  \widetilde{c}(\partial_\eta,M,m_t,\cos\theta,\mu) = {\rm Tr} \Big[
  \bm{H}(M,m_t,\cos\theta,\mu) \, \bm{\widetilde{s}}(\partial_\eta,M,m_t,\cos\theta,\mu)
  \Big] \, ,
\end{align}
where $\partial_\eta$ is a differential operator with respect to an auxiliary variable
$\eta$. The factorization formula for the hard-scattering kernel then takes the form
\cite{Ahrens:2010zv}
\begin{align}\label{eq:MasterFormula}
  C(z,M,m_t,\mu) = \int_{-1}^1 \! d\cos\theta \,\,
  \widetilde{c}(\partial_\eta,M,m_t,\cos\theta,\mu) \left( \frac{M}{\mu} \right)^{2\eta}
  \frac{e^{-2\gamma_E \eta}}{\Gamma(2\eta)} \frac{z^{-\eta}}{(1-z)^{1-2\eta}}
  \Bigg|_{\eta=0} .
\end{align}
To evaluate the above formula in terms of distributions in the variable $z$, one must use
analytic continuation to regulate the divergence at $z\to 1$, take derivatives with
respect to the auxiliary variable $\eta$, and then set $\eta \to 0$. This procedure yields
results in the form of plus distributions and delta functions given in (\ref{eq:Result})
below.

From the above discussion, we conclude that it is sufficient to focus on the calculation
of $\widetilde{c}$ at NNLO. Defining its perturbative expansion in analogy with
(\ref{eq:Cexp}), and using the expansions
\begin{equation}
  \begin{aligned}
    \bm{H} &= \alpha_s^2 \left[ {\bm H}^{(0)} + \frac{\alpha_s}{4\pi}\,{\bm H}^{(1)} +
      \left( \frac{\alpha_s}{4\pi} \right)^2 {\bm H}^{(2)} + \cdots \right] ,
    \\
    \bm{\widetilde{s}} &= \bm{\widetilde{s}}^{(0)} +
    \frac{\alpha_s}{4\pi}\,\bm{\widetilde{s}}^{(1)} + \left( \frac{\alpha_s}{4\pi}
    \right)^2 \bm{\widetilde s}^{(2)} + \cdots \, ,
  \end{aligned}
\end{equation}
for the hard and soft functions (\ref{eq:FF}), the result at NNLO reads
\begin{equation}\label{eq:TrExp}
  \begin{aligned}
    \widetilde{c}^{(2)}(\partial_\eta,M,m_t,\cos\theta,\mu) &= {\rm Tr} \left[
      \bm{H}^{(2)} \, \bm{\widetilde{s}}^{(0)} \right] + {\rm Tr} \left[ \bm{H}^{(0)} \,
      \bm{\widetilde{s}}^{(2)} \right] + {\rm Tr} \left[ \bm{H}^{(1)} \,
      \bm{\widetilde{s}}^{(1)} \right]
    \\
    &= \sum_{j=0}^4 \, c_j^{(2)}(M,m_t,\cos\theta,\mu) \, \partial_\eta^j \, .
  \end{aligned}
\end{equation}
In the second line we have defined expansion coefficients multiplying explicit powers of
$\partial_\eta$. We see that obtaining a complete result for the threshold expansion at
NNLO would amount to calculating the hard and soft matrices to this order. Although the
NLO coefficient $\widetilde{c}^{(1)}$ can be extracted from known results for the virtual
corrections and real emissions in the soft limit, the hard and soft matrices themselves
are so far unknown, except for in the absolute threshold limit $\rho \to 1$
\cite{Czakon:2009zw, Beneke:2009rj}. A main result of our work is the calculation of these
functions at NLO, for both the $q\bar{q}$ and $gg$ channels; details of the calculation
will appear in \cite{Ahrens:2010zv}. As for the NNLO corrections, we will show in what
follows that it is possible to extract all of the coefficients except for ${c}_0^{(2)}$ by
using the renormalization-group equations for the hard and soft functions, along with the
one-loop results and the two-loop anomalous dimensions from \cite{Becher:2009kw,
  Ferroglia:2009ep, Ferroglia:2009ii}. The renormalization group also determines the
$\mu$-dependent logarithms in this remaining term, but to obtain the scale-independent
piece would require to evaluate at NNLO the virtual corrections and soft real emissions.

We now sketch how to calculate the coefficient in (\ref{eq:TrExp}) using the evolution
equations for the hard and soft functions. The hard function satisfies
\cite{Kidonakis:1997gm}
\begin{align}\label{eq:Hev}
  \frac{d}{d\ln\mu} \, \bm{H}(M,m_t,\cos\theta,\mu) = \bm{\Gamma}_H \,
  \bm{H}(M,m_t,\cos\theta,\mu) + \bm{H}(M,m_t,\cos\theta,\mu) \, \bm{\Gamma}_H^\dagger \,
  ,
\end{align}
where $\bm{\Gamma}_H$ is given by the matrices $\bm{\Gamma}_{q\bar{q}}$ or
$\bm{\Gamma}_{gg}$ from \cite{Ferroglia:2009ii}. The anomalous dimensions to two-loop
order can be decomposed as
\begin{align}\label{eq:gammaH}
  \bm{\Gamma}_H = \Gamma_{\rm cusp}(\alpha_s) \left( \ln\frac{M^2}{\mu^2} - i\pi \right)
  \bm{1} + \bm{\gamma}^h(M,m_t,\cos\theta,\alpha_s) \, ,
\end{align}
where the object $\bm{\gamma}^h$ is defined through a comparison with the explicit results
of \cite{Ferroglia:2009ii}, and $\Gamma_{\rm cusp}$ is equal to $C_F\gamma_{\rm cusp}$ for
$q\bar{q}$ and $C_A\gamma_{\rm cusp}$ for $gg$, with $\gamma_{\rm cusp}$ the universal
cusp anomalous dimension. The evolution equation for the soft function follows from that
for the hard functions and PDFs, along with RG-invariance of the cross section. The result
is \cite{Ahrens:2010zv}
\begin{equation}\label{eq:Sev}
  \begin{aligned}
    &\frac{d}{d\ln\mu} \, \bm{\widetilde{s}}
    \left(\ln\frac{M^2}{\mu^2},M,m_t,\cos\theta,\mu\right)
    \\
    &= \bm{\Gamma}_s^\dagger\,\bm{\widetilde{s}}
    \left(\ln\frac{M^2}{\mu^2},M,m_t,\cos\theta,\mu\right) + \bm{\widetilde{s}}
    \left(\ln\frac{M^2}{\mu^2},M,m_t,\cos\theta,\mu\right) \bm{\Gamma}_s \, ,
  \end{aligned}
\end{equation}
where the soft anomalous dimension is given by
\begin{equation}\label{eq:gammaS}
  \begin{aligned}
    \bm{\Gamma}_s &= - \bigg[ \Gamma_{\rm cusp}(\alpha_s) \left( \ln\frac{M^2}{\mu^2} -
      i\pi \right) + 2\gamma^\phi(\alpha_s) \bigg] \, \bm{1} -
    \bm{\gamma}^h(M,m_t,\cos\theta,\alpha_s) \, .
  \end{aligned}
\end{equation}
The object $\gamma^\phi$ is defined through the large-$x$ limit of the Altarelli-Parisi
splitting functions, which reads
\begin{equation}
  P(x) = \frac{2\Gamma_{\rm cusp}(\alpha_s)}{(1-x)_+} +
  2\gamma^\phi(\alpha_s)\,\delta(1-x) \, .
\end{equation}
Since the anomalous dimensions for the hard function and PDFs are known to two-loop order,
we can calculate that of the soft function to this order.

Given the evolution equations, the anomalous dimensions, and the one-loop hard and soft
matrices, it is a simple matter to deduce a general expansion of the form
\begin{equation}\label{eq:Happrox}
  \begin{aligned}
    {\rm Tr} \left[ \bm{H}^{(n)}\,\bm{\widetilde s}^{(0)} \right] (M,m_t,\cos\theta,\mu)
    &= \sum_{j=0}^{2n} \, h_j^{(n)}(M,m_t,\cos\theta) \, \ln^j\frac{M^2}{\mu^2} \, ,
    \\
    {\rm Tr} \left[ \bm{H}^{(0)}\,\bm{\widetilde s}^{(n)} \right]
    (\partial_\eta,M,m_t,\cos\theta) &= \sum_{j=0}^{2n} \,
    s_j^{(n)}(M,m_t,\cos\theta)\,\partial_\eta^j \, ,
  \end{aligned}
\end{equation}
and to calculate all of the coefficients up to $h_0^{(2)}$ and $s_0^{(2)}$, which are left
undetermined. Along with the ${\rm Tr}[\bm{H}^{(1)}\bm{\widetilde{s}}^{(1)}]$
contributions, these determine $c_1^{(2)},\dots, c_4^{(2)}$ in (\ref{eq:TrExp}), as well
as the $\mu$-dependent piece of $c_0^{(2)}$.

The hard-scattering kernel $C$ is obtained by evaluating (\ref{eq:MasterFormula}). In
doing so, powers of $\partial_\eta$ in $\widetilde{c}$ are converted into delta functions
and plus distributions in the variable $z$. The final result can be written as
\begin{equation}\label{eq:Result}
  \begin{aligned}
    C^{(2)}(z,M,m_t,\mu) &= \int_{-1}^1 \! d\cos\theta \, \Bigg\{ D_3 \left[ \frac{\ln^3
        (1-z)}{1-z} \right]_+ + D_2 \left[ \frac{\ln^2(1-z)}{1-z} \right]_+
    \\
    &\quad\mbox{}+ D_1 \left[\frac{\ln(1-z)}{1-z}\right]_+ + D_0 \left[\frac{1}{1-z}
    \right]_+ + C_0\,\delta(1-z) + R(z) \Bigg\} \, .
  \end{aligned}
\end{equation}
Explicit results for the coefficients in both the $gg$ and $q\bar{q}$ channels as
functions of the variables $(M,m_t,\cos\theta,\mu)$ are given in the computer program
attached with the electronic version of this article. We stress that the coefficients
$D_0,\dots, D_3$ are completely determined from our calculations. The coefficient $C_0$,
on the other hand, is correct only in its $\mu$-dependence. It also receives
scale-independent contributions related to the translation from $\widetilde{c}$ to $C$,
and from the product of NLO hard and soft functions, which we discard for consistency. To
determine the remaining scale-independent piece would require a complete NNLO calculation
of the hard and soft matrices. Note that, because this piece is unknown, the expansion of
$C_0$ in terms of $\ln(M^2/\mu^2)$ considered so far is not unique: expansions in terms of
$\ln(m_t^2/\mu^2)$ (or other choices) still satisfy the same evolution equations. In our
phenomenological study in the next section we shall consider these two different
expansions as one indication of systematic errors related to the missing,
scale-independent piece of $C_0$. The function $R(z)$ is finite in the limit $z \to 1$ and
is not determined from the leading-order threshold expansion. However, it receives a
contribution related to the $z^{-\eta}$ factor in (\ref{eq:MasterFormula}). We choose to
keep these contributions when evaluating the NLO corrections, but drop them from the NNLO
ones. Our results for $D_1$, $D_2$, $D_3$, and the $\mu$-dependent terms agree with
\cite{Kidonakis:2003qe}, while the exact result for $D_0$ contains process-dependent
two-loop effects and is new.

The coefficients $D_i$ and $C_0$ do not depend on the variable $z$. It is therefore
possible to convolute the plus distributions and delta function with the luminosities as
in (\ref{eq:sigM}). For the phenomenologically interesting values of $0.03<\tau<0.3$
discussed in the next section, we can then make a few general comments concerning the
hierarchy of the different plus-distribution and delta-function terms. The contribution of
the leading $[\ln^3(1-z)/(1-z)]_+$ distribution is the largest in all cases. The
$[\ln^2(1-z)/(1-z)]_+$ distribution is the second largest, its contribution being about
1/3 that of the leading one. The contribution of the delta-function is about 10--20\% of
the leading plus distribution, and is of comparable size to the $[\ln(1-z)/(1-z)]_+$ and
$[1/(1-z)]_+$ distributions.

\section{Phenomenological applications}
\label{sec:pheno}

In this section we study the phenomenological implications of our results. The main goal
is to show the impact of the NNLO corrections on the central values and perturbative
uncertainties of the $t\bar{t}$ invariant mass distributions at the LHC and Tevatron.
However, given that our NNLO results are incomplete, it is important to clarify under what
conditions they are expected to give a good approximation to the full results. While it is
difficult to assign a systematic uncertainty to an unknown correction, we can examine in
parallel how well the same approximation works at NLO, and use this as an indication of
its expected validity at NNLO.

At NLO we can compare the following three calculations of the invariant mass distribution:
\begin{enumerate}
\item The full result, given by the Monte Carlo program MCFM \cite{Campbell:2000bg}.
\item The full threshold expansion, which is equivalent to the full result at leading
  order in $1-z$.
\item The approximate threshold expansion, which is equivalent to the full result at
  leading order in $1-z$ for the plus distributions and $\mu$-dependent terms, but not in
  the $\mu$-independent piece of the delta-function term. Here we also have the choice of
  writing the logarithms in the delta-function coefficients in terms of $\ln(m_t^2/\mu^2)$
  or $\ln(M^2/\mu^2)$.
\item The approximate threshold expansion, where the delta-function term is neglected
  entirely.
\end{enumerate}
To the extent that the second, third, and fourth options agree well with the full result
at NLO, it would seem plausible that our NNLO result, which is limited to the third and
fourth options, might also agree well with the full NNLO result, which is unknown. We thus
investigate these different approximations to the invariant mass distributions for a range
of $M$ and $\mu$, and study under what circumstances they show a reasonable agreement. The
naive expectation would be that they agree only at very high values of the invariant mass,
since then $\tau=M^2/s \to 1$ and the integrand in (\ref{eq:sigM}) is needed only in the
$z \to 1$ limit. Moreover, the integrals over the plus distributions in (\ref{eq:Result})
are more singular at $\tau \to 1$ than the unknown portion of the correction proportional
to the delta function. However, due to the rapid fall-off of the PDFs at large $x$, the
distribution at very high $\tau$ is difficult to measure. The most interesting region for
phenomenology would be from $M \sim 2m_t$ to around 1 TeV at the Tevatron and up to
several TeV at the LHC; this region of the invariant mass corresponds to $\tau<0.3$. In
order for our results to be a good approximation in this region, it is necessary that the
luminosity functions $\ff_{ij}(\tau/z,\mu)$ fall off so fast for $\tau/z\to 1$ that only
the largest values of $z$ give significant contributions to the integrand in
(\ref{eq:sigM}). In that case, a dynamical enhancement of the partonic threshold region
may occur, even if $\tau$ is not close to unity \cite{Becher:2007ty, Ahrens:2008qu,
  Ahrens:2008nc}. The agreement of the three different approximations at moderate values
of $\tau$ is thus a measure of whether this dynamical enhancement actually occurs.

\begin{figure}[t!]
  \begin{center}
    \begin{tabular}{lr}
      \includegraphics[width=0.40\textwidth]{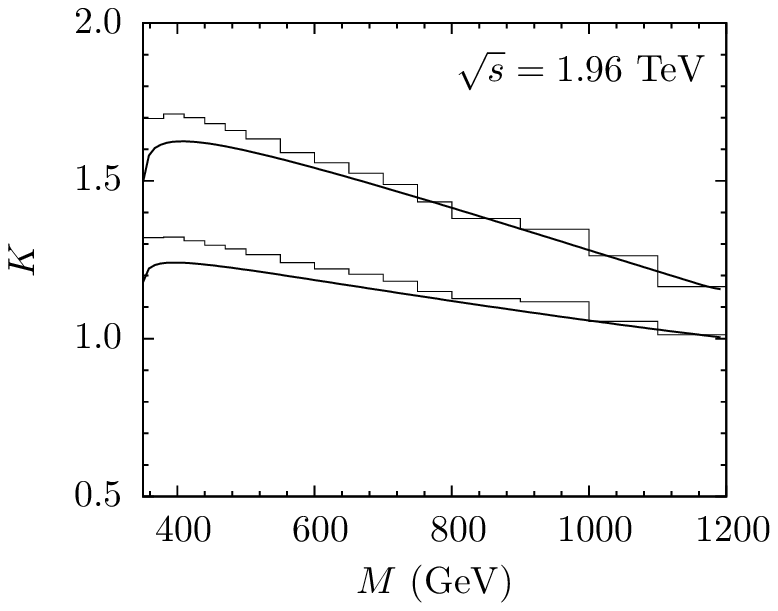}
      &
      \includegraphics[width=0.40\textwidth]{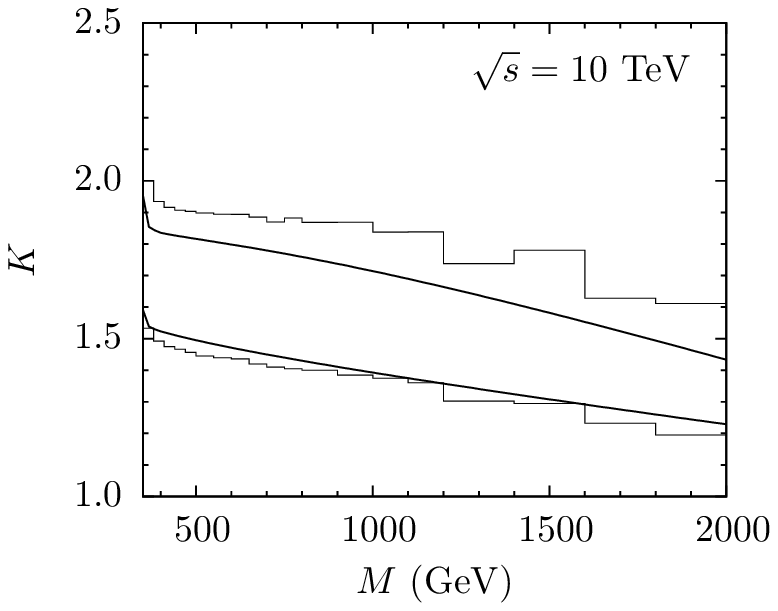}
    \end{tabular}
  \end{center}
  \vspace{-0.5cm}
  \caption{\label{fig:MCFM} Ratio $K=(d\sigma/dM)/(d\sigma^{{\rm LO}}(\mu=2m_t)/dM)$
    versus the $t\bar t$ invariant mass $M$ for the Tevatron (left) and the LHC with
    $\sqrt{s}=10$\,TeV (right). The binned histograms show the exact NLO results from the
    Monte Carlo program MCFM \cite{Campbell:2000bg}, while the solid curves are obtained
    from the threshold expansion at NLO. In each case the upper lines correspond to
    $\mu=m_t$ and the lower ones to $\mu=4m_t$.}
\end{figure}

\begin{figure}[t!]
  \begin{center}
    \begin{tabular}{lr}
      \psfrag{left}[][][0.7]{$d\sigma/dM$ (fb/GeV)}
      \psfrag{MU}[][][0.7]{$\mu$ (GeV)}
      \psfrag{S}[][][0.7]{$\sqrt{s}=1.96$~TeV}
      \psfrag{M}[][][0.7]{$M=400$~GeV}
      \includegraphics[width=0.38\textwidth]{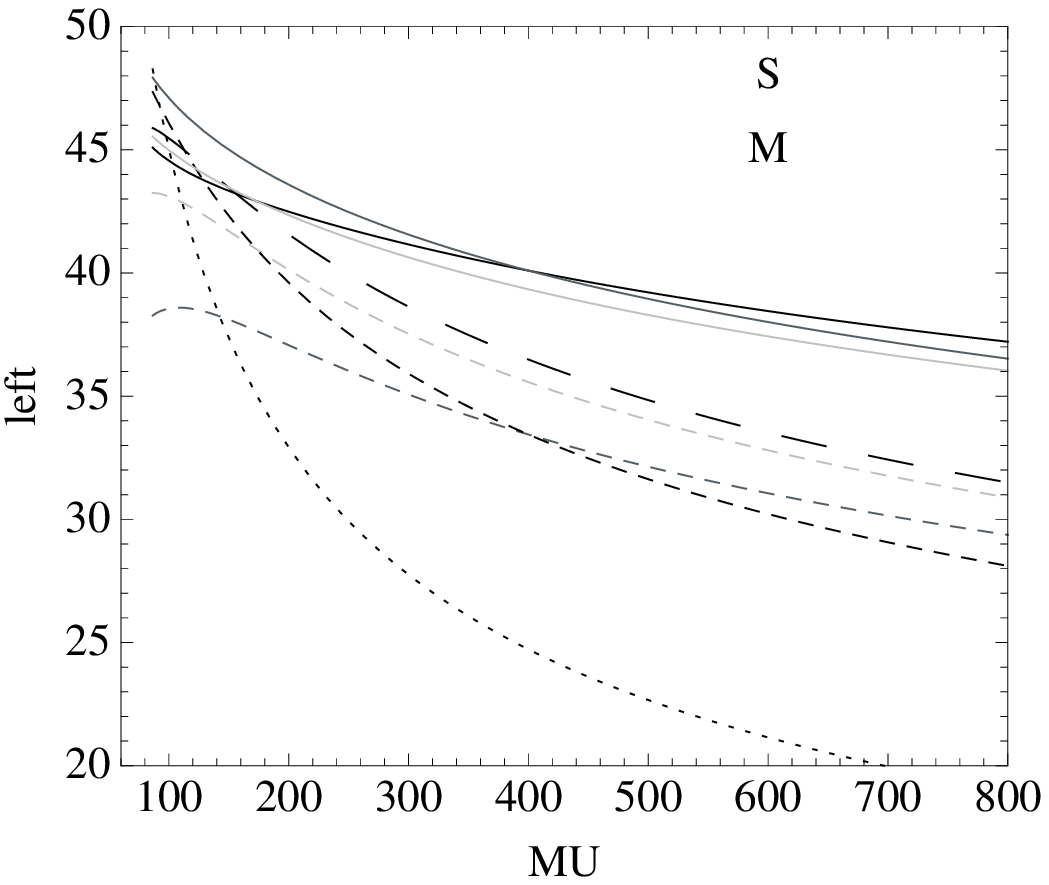}
      &
      \psfrag{left}[][][0.7]{$d\sigma/dM$ (fb/GeV)}
      \psfrag{MU}[][][0.7]{$\mu$ (GeV)}
      \psfrag{S}[][][0.7]{$\sqrt{s}=1.96$~TeV}
      \psfrag{M}[][][0.7]{$M=1000$~GeV}
      \includegraphics[width=0.38\textwidth]{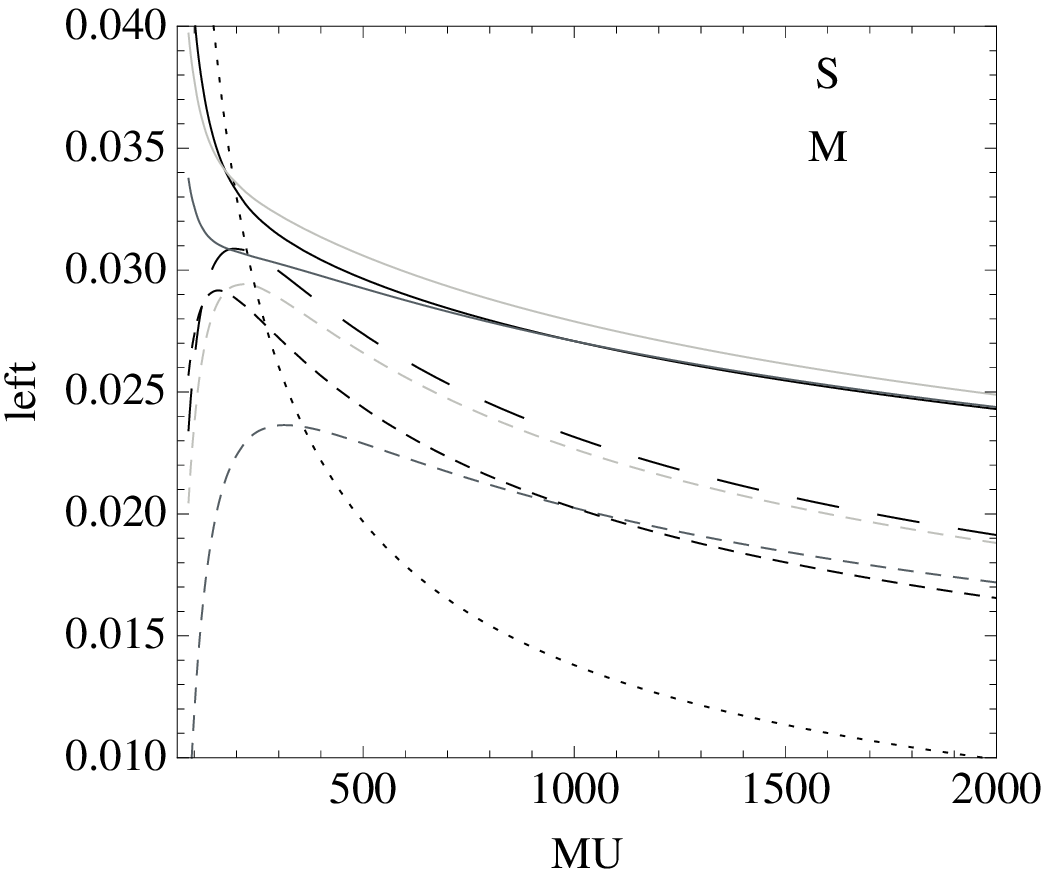}
      \\
      &
      \\
      \psfrag{left}[][][0.7]{$d\sigma/dM$ (fb/GeV)}
      \psfrag{MU}[][][0.7]{$\mu$~(GeV)}
      \psfrag{S}[][][0.7]{$\sqrt{s}=10$~TeV}
      \psfrag{M}[][][0.7]{$M=400$~GeV}
      \includegraphics[width=0.38\textwidth]{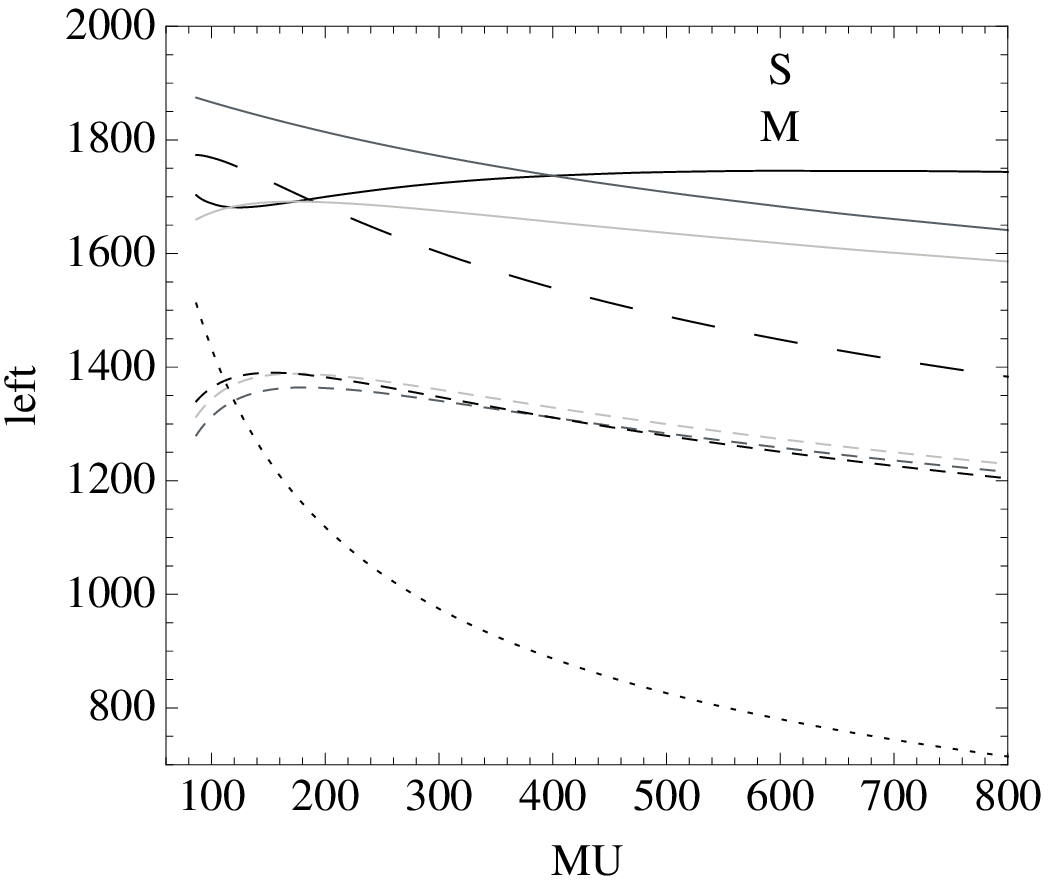}
      &
      \psfrag{left}[][][0.7]{$d\sigma/dM$ (fb/GeV)}
      \psfrag{MU}[][][0.7]{$\mu$~(GeV)}
      \psfrag{S}[][][0.7]{$\sqrt{s}=10$~TeV}
      \psfrag{M}[][][0.7]{$M=1000$~GeV}
      \includegraphics[width=0.38\textwidth]{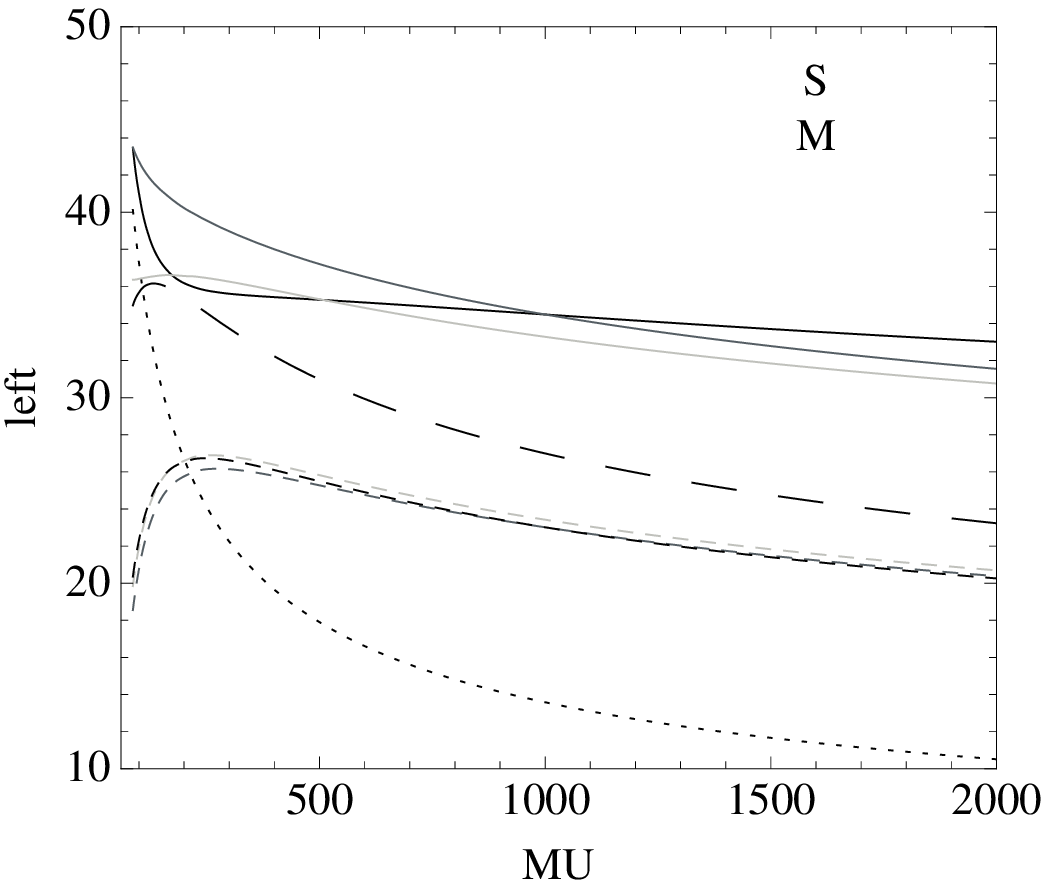}
    \end{tabular}
  \end{center}
  \vspace{-0.5cm}
  \caption{\label{fig:ScaleVarM} Scale dependence of the $t\bar{t}$ invariant mass
    distribution at the Tevatron (upper plots) and LHC (lower plots), for two different
    values of $M$. The dotted, dashed, and solid lines show results obtained using the
    threshold expansion at LO, NLO, and NNLO, respectively. The long-dashed black lines
    are obtained using the full threshold expansion at NLO. The light-grey (dark-grey)
    lines correspond to the approximate threshold expansion (option~3), in which the
    logarithms in the delta-function coefficients are written in terms of
    $\ln(m_t^2/\mu^2)$ ($\ln(M^2/\mu^2)$), and the black lines to the threshold expansion
    where the delta-function term is dropped altogether (option 4).}
\end{figure}

\begin{figure}[t!]
  \begin{center}
    \begin{tabular}{lr}
      \includegraphics[width=0.40\textwidth]{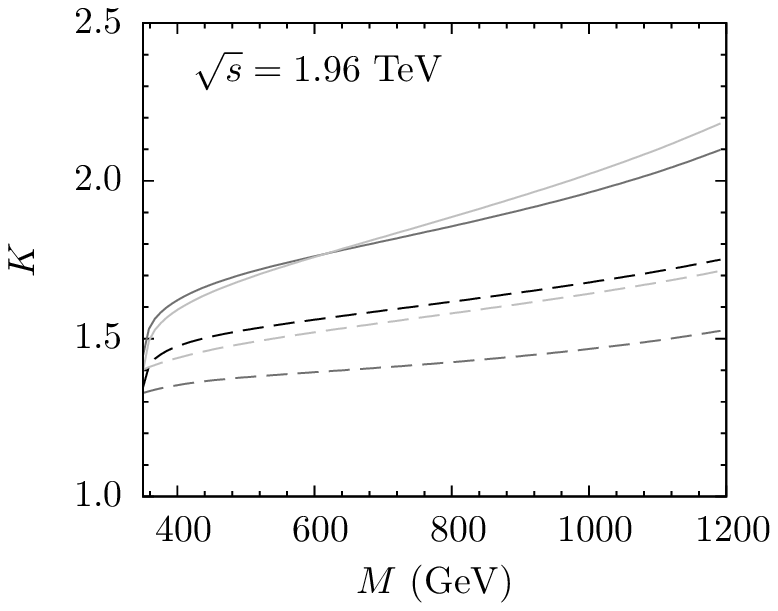}
      &
      \includegraphics[width=0.40\textwidth]{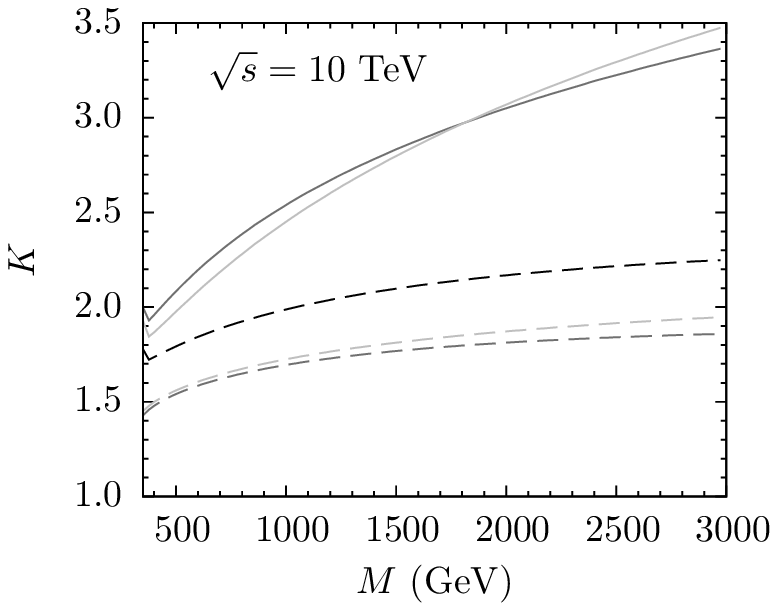}
    \end{tabular}
  \end{center}
  \vspace{-0.5cm}
  \caption{\label{fig:NLOratiosM} Ratio $K=(d\sigma/dM)/(d\sigma^{{\rm LO}}(\mu=M)/dM)$
    versus the $t\bar t$ invariant mass $M$ for the Tevatron (left) and LHC (right), for
    the default scale choice $\mu=M$. The dashed and solid lines show results obtained
    using the threshold expansion at NLO and NNLO, respectively. The black-dashed lines
    are obtained using the full threshold expansion at NLO. The light-grey (dark-grey)
    lines correspond to the approximate threshold expansion (option~3), in which the
    logarithms in the delta-function coefficients are written in terms of
    $\ln(m_t^2/\mu^2)$ ($\ln(M^2/\mu^2)$).}
\end{figure}

\begin{figure}[t!]
  \begin{center}
    \begin{tabular}{lr}
      \includegraphics[width=0.40\textwidth]{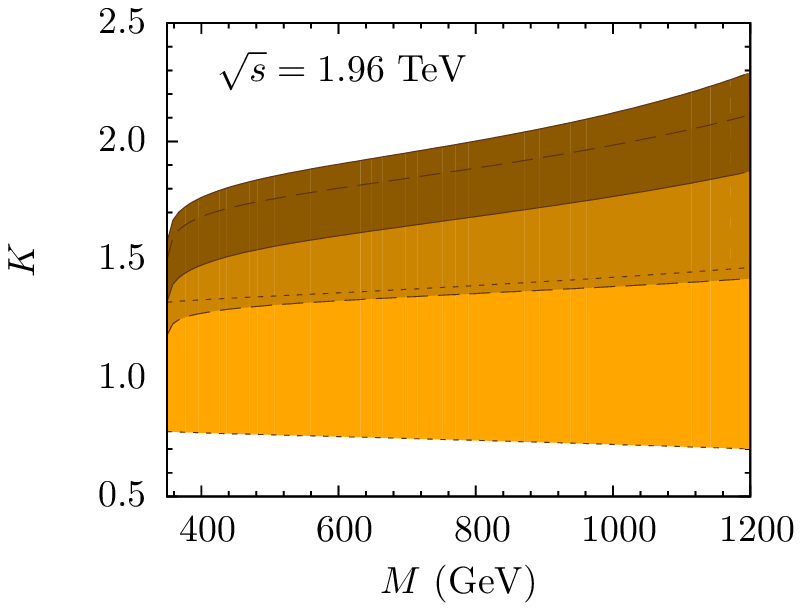}
      &
      \includegraphics[width=0.40\textwidth]{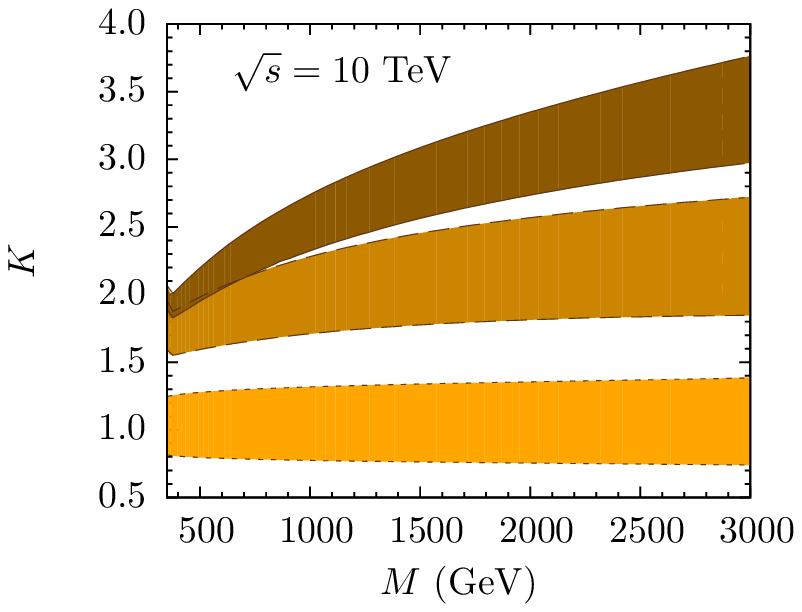}
      \\
      & 
      \\
      \includegraphics[width=0.40\textwidth]{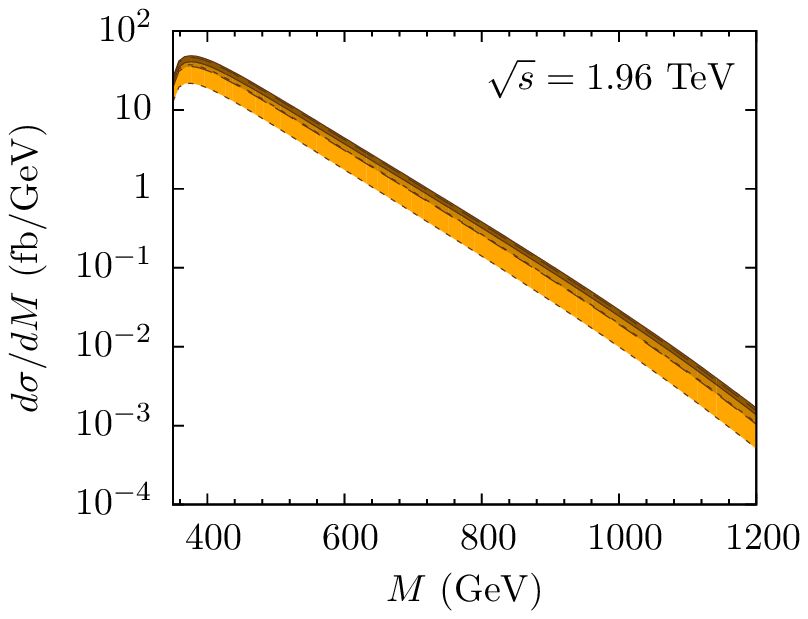}
      &
      \includegraphics[width=0.40\textwidth]{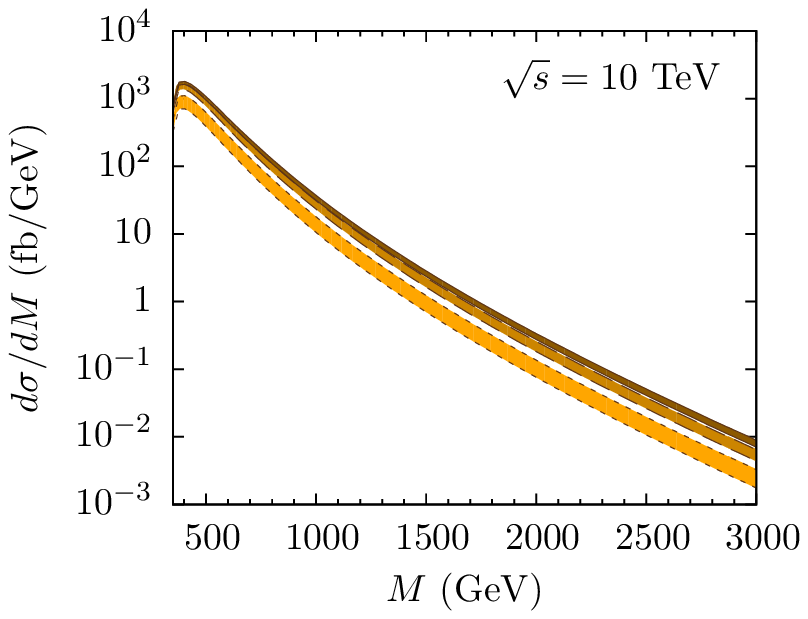}
    \end{tabular}
  \end{center}
  \vspace{-0.5cm}
  \caption{\label{fig:Kbands} Top: Ratio $K=(d\sigma/dM)/(d\sigma^{{\rm LO}}(\mu=M)/dM)$
    versus the $t\bar t$ invariant mass for the Tevatron (left) and LHC (right), with the
    factorization scale varied in the range $M/2<\mu<2M$. The light bands between the
    dotted lines correspond to the threshold expansion at LO, the medium bands between the
    dashed lines to NLO, and the dark bands between the solid lines to NNLO. Bottom: Same
    as above, but for the invariant mass distribution $d\sigma/dM$. }
\end{figure}

With these points in mind, we begin by comparing the exact result and the complete
threshold expansion at NLO. The results are shown in Figure~\ref{fig:MCFM}. Here and
throughout the analysis we use the MSTW2008 NNLO PDFs \cite{Martin:2009bu}, take
$\alpha_s(M_Z)=0.117$ with three-loop running in the $\overline{\rm MS}$ scheme with five
active flavors, and use $m_t=173.1$~GeV in the pole scheme. We use the same set of PDFs at
each order of perturbation theory so as to better illustrate the size of the different
perturbative contributions to the hard-scattering kernels $C_{ij}$. We choose $\mu=2m_t$
as the central value and vary it between $m_t$ and $4m_t$. For the case of the Tevatron,
we see that the threshold expansion is in good agreement with the full result even at
values $M \sim 2m_t$, where $\tau \approx 0.03$, and gets progressively better at higher
values of $M$. The situation is only slightly worse for the LHC with $\sqrt{s}=10$~TeV,
even though the value of $\tau$ for a given $M$ is about 25 times smaller. We also note
that at LHC energies the threshold expansion matches the exact result better at the higher
value of $\mu=4m_t$ than at $\mu=m_t$. In what follows, we shall also explore the scale
choice $\mu \sim M$. Since such a choice is not practical in the Monte Carlo program MCFM,
from now on we shall use the exact threshold expansion to represent the full NLO result,
keeping in mind the good agreement seen in Figure~\ref{fig:MCFM}.

We next investigate in more detail the approximate threshold expansions (options~3 and 4
above) at NLO and NNLO. In Figures~\ref{fig:ScaleVarM} and \ref{fig:NLOratiosM} we compare
the full threshold expansion at NLO with the different approximate expansions at the same
order, and also show the approximate expansions at NNLO (these are generated by adding the
NNLO corrections to the full threshold expansion at NLO). In Figure~\ref{fig:ScaleVarM} we
show the differential cross section as a function of $\mu$ for two different values of
$M$. We notice that the different approximations at NLO show better agreement with the
full threshold expansion for higher values $\mu \sim M$, especially at LHC energies, where
the NLO approximations at lower values of $\mu$ differ greatly from the full results. At
NNLO, the different approximations in options~3 and 4 do not differ substantially from one
another, when one considers their highest and lowest values in the range $M/2 < \mu < 2M$,
which is how perturbative uncertainties are typically estimated. In
Figure~\ref{fig:NLOratiosM} we show results for the same approximations at NLO and NNLO,
but this time as a function of $M$. Given the better agreement at higher $\mu$ observed in
the previous two figures, we have made the default choice $\mu=M$. We observe that at NLO
the approximate threshold expansions recover a significant portion of the exact NLO
correction, both at the Tevatron and LHC. It is not unreasonable to expect that the same
is true of our approximate NNLO results, although they are clearly no substitute for a
complete NNLO computation.

In Figure~\ref{fig:Kbands} we show the invariant mass spectrum as a function of $M$, with
bands representing the uncertainty associated with scale variations in the range
$M/2<\mu<2M$. For the NNLO result, we have displayed only the approximate threshold
expansion in which the logarithms in the delta-function coefficient are written in terms
of $\ln(M^2/\mu^2)$; those in the other two NNLO approximations are similar, and very
nearly contained within this band. At both the Tevatron and the LHC we note a reduction of
scale uncertainty upon including the NNLO corrections, which is the expected benefit of
including more terms in the perturbative expansion. However, we must also observe from
Figure~\ref{fig:ScaleVarM} that at NLO the perturbative uncertainty estimated by varying
the scale in the range $M<\mu<2M$ in the approximate threshold expansions is actually
smaller than in the complete threshold expansion at both $M=400$~GeV and $M=1000$~GeV at
the LHC. We thus cannot rule out that the very small uncertainty of the NNLO result at
small values of $M$ is also an underestimate.

\section{Conclusions}

We calculated $\mathcal{O}(\alpha_s^4)$ contributions to the $t\bar{t}$ invariant mass
distribution at hadron colliders. The calculation was based on an effective field-theory
approach to the factorization of the perturbative hard-scattering kernels into
matrix-valued hard and soft functions in the partonic threshold region $\hat{s} \gtrsim
M^2$. At the technical level, it involved calculating the hard and soft functions at NLO,
and then determining at NNLO the logarithmic $\mu$-dependence of these functions using the
renormalization group and recent results for two-loop anomalous dimensions. Our
computations yield exact results for the coefficients of all singular plus distributions
and $\mu$-dependent logarithms in the differential partonic cross section at NNLO, in the
limit where the invariant mass of the $t\bar{t}$ pair approaches the partonic
center-of-mass energy. They agree with those previously obtained in
\cite{Kidonakis:2003qe} using a Mellin-space approach to soft gluon resummation, but go
beyond them by uniquely determining the coefficient of the $[1/(1-z)]_+$ distribution,
which contains process-dependent two-loop effects related to anomalous dimensions of the
soft and hard functions. To obtain the missing scale-independent piece of the
delta-function term would require the exact two-loop calculation of these functions.

In Section \ref{sec:pheno} we performed a numerical study of the $t\bar{t}$ invariant mass
distribution at Tevatron and LHC energies. In both cases, we noted the expected decrease
in scale uncertainty after including the NNLO corrections, and also noticeable changes in
the central values, especially at higher values of the invariant mass. Given that our NNLO
result is still incomplete, even in the threshold region, part of our analysis was focused
on determining how well the analogous approximation works at NLO. We observed that at NLO
the full result in QCD and its exact threshold expansion agree fairly well. The same was
true of the approximate threshold expansion at NLO, where the scale-independent pieces of
the delta-function term are neglected. This provides some evidence that the steep fall-off
of the PDFs at large $x$ induces a dynamical enhancement of the partonic threshold region,
which would mean that the terms we have calculated at NNLO provide a useful approximation
to the full result.

Our finding that the NNLO perturbative corrections, which we have calculated in this
Letter, have quite an important impact in the region of large invariant masses, as seen in
Figure~\ref{fig:Kbands}, provides a strong motivation for improving this estimate with
more sophisticated calculations. We expect that the scale dependence indicated by the
width of the various bands in the figure can be reduced significantly by resumming the
partonic threshold logarithms to all orders in perturbation theory. Based on the results
presented here and in \cite{Ferroglia:2009ep,Ferroglia:2009ii}, such a resummation is now
possible at NNLL order. It will be discussed in a forthcoming paper \cite{Ahrens:2010zv}.
Ultimately, however, only a complete NNLO calculation of the spectrum will provide us with
a complete picture of the true residual uncertainty in the theoretical predictions.

\end{document}